\RequirePackage{fixltx2e}
\documentclass[aip,jcp,reprint,showpacs]{revtex4-1}
\usepackage{amsmath,amsfonts} 
\usepackage{wasysym}
\usepackage[acronym]{glossaries}
\usepackage{graphicx}
\usepackage{dcolumn} 
\usepackage{bm} 
\usepackage{ifpdf} 
\usepackage{natbib}
\usepackage[byname]{smartref}
\usepackage[pdfusetitle, %
breaklinks, %
colorlinks, %
linkcolor=blue, %
citecolor=blue, %
urlcolor=blue]{hyperref}
\usepackage[table]{xcolor}
\usepackage{multirow}
\usepackage{braket}

\makeatletter %
\hypersetup { %
  pdftitle={\@title}, %
  pdfauthor={\@author}, %
  pdfsubject={}, %
  pdfkeywords={} %
} %
\makeatother

\newcommand{\bfr}{\mathbf{r}}
\newcommand{\bfR}{\mathbf{R}}
\newcommand{\SB}{\mathbf{s}}
\newcommand{\eq}[1]{Eq.~(\ref{#1})}
\newcommand{\eqs}[1]{Eqs.~(\ref{#1})}
\newcommand{\fig}[1]{Fig.~\ref{#1}}

\newacronym{CI}{CI}{conical intersection} %
\newacronym{GP}{GP}{geometric phase} %
\newacronym{LVC}{LVC}{linear vibronic coupling} %
\newacronym{DOF}{DOF}{degrees of freedom} %
\newacronym{PES}{PES}{potential energy surface} %
\newacronym{QCL}{QCL}{quantum-classical Liouville} %
\newacronym{WT}{WT}{Wigner transform} %

\begin{document}
\title{Analysis of geometric phase effects in the quantum-classical Liouville formalism}

\author{Ilya G. Ryabinkin} %
\affiliation{Department of Physical and Environmental Sciences,
  University of Toronto Scarborough, Toronto, Ontario M1C 1A4,
  Canada} %
\affiliation{Chemical Physics Theory Group, Department of Chemistry,
  University of Toronto, Toronto, Ontario M5S 3H6, Canada} %
\author{Chang-Yu Hsieh} %
\author{Raymond Kapral} %
\affiliation{Chemical Physics Theory Group, Department of Chemistry,
  University of Toronto, Toronto, Ontario M5S 3H6, Canada} %
\author{Artur F. Izmaylov} %
\affiliation{Department of Physical and Environmental Sciences,
  University of Toronto Scarborough, Toronto, Ontario M1C 1A4,
  Canada} %
\affiliation{Chemical Physics Theory Group, Department of Chemistry,
  University of Toronto, Toronto, Ontario M5S 3H6, Canada} %

\date{\today}

\begin{abstract}
  We analyze two approaches to the quantum-classical Liouville (QCL)
  formalism that differ in the order of two operations: Wigner
  transformation and projection onto adiabatic electronic states.  The
  analysis is carried out on a two-dimensional linear vibronic model
  where geometric phase (GP) effects arising from a conical
  intersection profoundly affect nuclear dynamics. We find that the
  Wigner-then-Adiabatic (WA) QCL approach captures GP
  effects, whereas the Adiabatic-then-Wigner (AW) QCL approach does not.
  Moreover, the Wigner transform in AW-QCL leads to an ill-defined
  Fourier transform of double-valued functions. The double-valued
  character of these functions stems from the nontrivial GP of adiabatic
  electronic states in the presence of a conical intersection. In
  contrast, WA-QCL avoids this issue by starting with the Wigner
  transform of single-valued quantities of the full problem.  Since
  the WA-QCL approach uses solely the adiabatic potentials and
  non-adiabatic derivative couplings as an input, our results indicate
  that WA-QCL can capture GP effects in general two-state crossing
  problems using first-principles electronic structure calculations
  without prior diabatization or introduction of explicit phase
  factors.
\end{abstract}


\maketitle

\section{Introduction}
\label{sec:introduction}

Molecular electronic adiabatic surfaces often cross forming degenerate
manifolds of nuclear configurations with the topology of
\glspl{CI}.\cite{Migani:2004/271,Yarkony:1996/rmp/985} \glspl{CI} are
the most common triggers of radiationless transitions that drive
photo-induced chemistry,\cite{Migani:2004/271} and transfers of electronic energy and charge.
\cite{Blancafort:2005/JACS/3391,Blancafort:2001/JACS/722, Izmaylov:2011/jcp/234106} 
Besides facilitating electronic transitions \glspl{CI} change the topology of
the nuclear subspace: if adiabatic electronic wave functions are
infinitely slowly (adiabatically) transported around a closed loop
that encircles the \gls{CI} seam, they acquire an extra $(-1)$ phase.
This is the \gls{GP} that makes the electronic wave functions
double-valued functions of nuclear
coordinates.\cite{LonguetHigg:1958/rspa/1, Berry:1984/rspa/45,
  Simon:1983/prl/2167, Ham:1987/prl/725, Berry:1987/rspa/31,
  Aitchison:1988/ps/12} In order to have a single-valued total
electron-nuclear wave function, the nuclear wave functions have to
compensate for sign changes in their electronic counterparts; hence, a
nuclear Schr\"odinger equation must be solved with double-valued
boundary conditions.  The extra phase causes interference between
parts of the nuclear wave function, which results in different
vibronic spectra and nuclear dynamics as compared to the case without
explicit account for the \gls{GP}.~\cite{LonguetHigg:1958/rspa/1, Schon:1994/cpl/55,
  Schon:1995/jcp/9292, Ryabinkin:2013/prl/220406, Loic:2013un}

The double-valued character of the nuclear wave functions in the
adiabatic representation is challenging for numerical
simulations. Switching to the diabatic
representation\cite{Cederbaum:2004/CI} makes the nuclear diabatic wave
function single-valued.  However, the diabatic representation within a
finite electronic subspace cannot be obtained~\cite{Baer:1975/cpl/112}
for a general polyatomic system ($N_\text{atoms} > 2$), so one has to
resort to approximate diabatization
schemes.~\cite{Koppel:2006/mp/1069,Voorhis:2010/arpc/149,Sirjoosingh:2011/jctc/2831,
  Yarkony:2012/cr/481,Subotnik:2009/jcp/234102,Opalka:2010/jcp/154108,
  Troisi:2003/jcp/5356,Papas:2008/jcp/124104,Nakamura:2001/jcp/10353}
To remain in the unambiguous adiabatic representation,
\citet*{Mead:1979/jcp/2284} proposed to compensate for the \gls{GP} of
individual nuclear states by attaching an extra phase factor
$e^{i\lambda(\mathbf{R})}$, where $\lambda(\mathbf{R})$ is a function
that increases by $\pi$ on encircling a closed path around the
\gls{CI} seam. This approach was successfully
implemented~\cite{Kendrick:1995/jcp/4160, Kendrick:2002/cp/31,
  Kendrick:2003} and applied to many real molecules by
Kendrick.\cite{Kendrick:1996/jcp/7475, Kendrick:1996/jcp/7502,
  Kendrick:1997/prl/2431, Kendrick:2003/jpca/6739} However, the
definition of $\lambda(\mathbf{R})$ is based on diabatic model
considerations, and its application to a general molecular system
assumes some diabatic model potential. An alternative approach to the
definition of $\lambda(\mathbf{R})$ was suggested by Baer and coworkers.
\cite{Baer:1997/jcp/2694, Xu:2000/jcp/2746} They related
$\nabla\lambda(\mathbf{R})$ to the matrix elements of derivative
couplings $\mathbf{d}_{12}^{(1)}(\mathbf{R})$, thus removing any
arbitrariness in the definition of $\lambda(\mathbf{R})$. However, this
suggestion was challenged by~\citet{Kendrick:1999/jcp/7594} who 
found that Baer's approach inconsistently
neglects some terms, and this
inconsistency leads one to question the validity of the final
result. Therefore, accounting for \gls{GP} effects for general systems
using only results of electronic structure calculations seems to be challenging within a fully
quantum approach.

In this context, it is interesting that the \gls{QCL}
formalism~\cite{Kapral:2006kw} is able to capture
\gls{GP} effects~\cite{Kelly:2010/jcp/084502} using only adiabatic
input: energies and non-adiabatic couplings.  The \gls{QCL} framework
uses the \gls{WT} of the nuclear \gls{DOF} to arrive at a mixed
quantum-classical description.  Derivations of the \gls{QCL} equations
for non-adiabatic problems using the adiabatic representation for
electronic \gls{DOF} can proceed along two paths that differ in the
order in which the \gls{WT} and the projection to the adiabatic basis
are applied.  We will denote by WA-QCL the approach where the \gls{WT}
is done first,\cite{Kapral:1999/jcp/8919} and by AW-QCL the approach where the adiabatic
projection precedes the \gls{WT}.\cite{Horenko:2002/jcp/11075,Ando:2003/jcp/10399}  
It is not {\it a priori} obvious which of the two approaches is better.  Previously, the WA-QCL and
AW-QCL approaches were used to simulate the spin-boson model and no significant
differences were found.\cite{Ando:2003/jcp/10399} In this work we
analyze the efficacy of these approaches to describe the dynamics in a
two-dimensional (2D) \gls{LVC} model.  As has been shown
recently~\cite{Ryabinkin:2013/prl/220406}, the 2D \gls{LVC} model
exhibits prominent \gls{GP} effects, such as a distinct interference
pattern due to the \gls{GP} as well as a strong impact of the \gls{GP}
on nuclear dynamics. This model allows us to assess both \gls{QCL}
approaches with respect to CI-introduced topological features in
non-adiabatic dynamics.

The paper is organized as follows: first, we briefly illustrate the
emergence of double-valued adiabatic electronic wave functions in the
2D \gls{LVC} model. Then, we outline the derivations of the matrix form of the WA-QCL
equation starting from the full electron-nuclear density
matrix~\cite{Kapral:1999/jcp/8919} and the AW-QCL equation starting
from a projection of the Liouville equation onto an adiabatic
electronic basis.~\cite{Horenko:2002/jcp/11075,
  Ando:2003/jcp/10399} Comparing these two approaches we identify a
term which is responsible for \gls{GP} effects. We conclude our paper
with numerical results and their analysis.

\section{Geometric phase in a two-dimensional linear vibronic coupling
  model}
\label{sec:geometric-phase-two}

The two-dimensional \gls{LVC} is a prototype for systems containing a
\gls{CI} and exhibiting nontrivial \gls{GP} effects. The
electron-nuclear Hamiltonian of the model is
\begin{equation}
  \label{eq:H_lvc}
  \hat{\mathbf{H}} = {\hat T}_N  \mathbf{1}_2 + 
  \begin{pmatrix} 
    \hat V_{11} & \hat V_{12} \\
    \hat V_{12} & \hat V_{22}
  \end{pmatrix},
\end{equation}
where ${\hat T}_N=-\frac{\hbar^2}{2} (\partial^2 /\partial x^2
+\partial^2 /\partial y^2) $ is the nuclear kinetic energy operator,
$\hat V_{11}$ and $\hat V_{22}$ are the diabatic potentials
represented by identical 2D parabolas shifted in the $x$-direction by $a$
and coupled by the $\hat V_{12}$ potential:
\begin{align}
  \label{eq:diab-me-11}
  {\hat V}_{11} = {} & \dfrac{\omega^2}{2}\left[\left(x +
      \dfrac{a}{2}\right)^2
    + y^2\right], \quad  {\hat V}_{12} = c y, \\
  \label{eq:diab-me-22}
  {\hat V}_{22} = {} & \dfrac{\omega^2}{2}\left[\left(x -
      \dfrac{a}{2}\right)^2 + y^2\right].
\end{align}
Electronic \gls{DOF} are represented as position-independent diabatic
states $\ket{1}$ and $\ket{2}$ in a two-dimensional electronic
subspace.

Transformation to the adiabatic representation is made by diagonalizing 
the potential matrix in Eq.~(\ref{eq:H_lvc}) by means of the unitary
transformation $U$
\begin{equation}
  \label{eq:U-mat}
  \mathbf{U}(\theta) = 
  \begin{pmatrix}
    \cos\frac{\theta}{2} & -\sin\frac{\theta}{2} \\[1ex]
    \sin\frac{\theta}{2} & \cos\frac{\theta}{2}
  \end{pmatrix},
\end{equation}
where $\theta$ is a mixing angle between the diabatic states, and the
factor $\frac{1}{2}$ is introduced for convenience. The adiabatic
electronic wave functions are related to the columns of $U$ as
\begin{align}
  \label{eq:psi_el_ad_1}
  \ket{\psi_1^\text{adi}} = {} & \cos\frac{\theta}{2}\ket{1} +
  \sin\frac{\theta}{2}\ket{2}   \\
  \label{eq:psi_el_ad_2}
  \ket{\psi_2^\text{adi}} = {} & -\sin\frac{\theta}{2}\ket{1} +
  \cos\frac{\theta}{2}\ket{2}.
\end{align}
The angle $\theta$ is determined by the matrix elements of the
diabatic Hamiltonian~(\ref{eq:H_lvc}) as
\begin{equation}
  \label{eq:theta}
  \theta = \arctan \dfrac{2\,V_{12}}{V_{11} - V_{22}},
\end{equation}
and changes by $2\pi$ for any closed path in the nuclear $(x,y)$
subspace that encircles the \gls{CI} point. Considering that $U(2\pi)
=-\mathbf{1}_2$, both adiabatic wave functions
$\ket{\psi_i^\text{adi}}$ acquire an extra $(-)$ sign, which is a
consequence of the \gls{GP}. To have a single-valued total wave
function
\begin{equation}
  \label{eq:Psi_sep}
  \ket{\Psi} =
  \ket{\chi_1^\text{adi}}\ket{\psi_1^\text{adi}} +
  \ket{\chi_2^\text{adi}}\ket{\psi_2^\text{adi}},  
\end{equation}
a sign change must be also imposed on the adiabatic nuclear wave
functions $\ket{\chi_{1,2}^\text{adi}}$.

This consideration can be extended to the total density operator
$\hat\rho = \ket{\Psi}\bra{\Psi}$, which is also a single-valued
function of the nuclear $(x,y)$ coordinates.  In contrast, the nuclear
density matrix in the adiabatic representation
\begin{eqnarray}
  \label{eq:nuc-den-adi}
  \hat X_{\alpha\alpha'} =
  \Braket{\psi_{\alpha}^\text{adi}|{\hat\rho}|\psi_{\alpha'}^\text{adi}}
  = \ket{\chi_{\alpha}^\text{adi}}\bra{\chi_{\alpha'}^\text{adi}},   
\end{eqnarray}
is a complicated object due to the double-value boundary conditions
imposed on $\ket{\chi_{\alpha}^\text{adi}}$ and
$\bra{\chi_{\alpha'}^\text{adi}}$.

To model dynamics of the nuclear adiabatic densities
$\hat{\mathbf{X}}$ we solve the corresponding Liouville equation
\begin{equation}
  \label{eq:tdse-nuc}
  i\hbar\frac{\partial \hat{\mathbf{X}}}{\partial t} 
  = [\hat{\mathbf{H}}^\text{adi},\hat{\mathbf{X}}].
\end{equation}
Here, the Hamiltonian $\hat{\mathbf{H}}^\text{adi}$ is obtained by
projecting the electronic \gls{DOF} in Hamiltonian~(\ref{eq:H_lvc})
onto the $\ket{\psi_{1,2}^\text{adi}}$ subspace:
\begin{equation}
  \label{eq:Hadi}
  \hat{\mathbf{H}}^\text{adi} = 
  \begin{pmatrix}
    \hat T_N + \hat{\boldsymbol\tau}_{11} + {\hat W}_{-} &  \hat{\boldsymbol\tau}_{12} \\
    \hat{\boldsymbol\tau}_{21} & \hat T_N + \hat{\boldsymbol\tau}_{22}
    + {\hat W}_{+}
  \end{pmatrix},
\end{equation}
where
\begin{equation}
  \label{eq:NAC-def}
  \begin{split}
    \hat{\boldsymbol\tau}_{ij} = {} &
    -\hbar^2[\langle\psi_i^\text{adi} | \nabla\psi_j^\text{adi}\rangle
    \cdot \nabla + \frac{1}{2}\langle\psi_i^\text{adi} |
    \nabla^2\psi_j^\text{adi}\rangle] \\
    {} \equiv {} & -\hbar^2[\mathbf{d}_{ij}^{(1)}\cdot\nabla +
    \frac{1}{2} d_{ij}^{(2)}].
  \end{split}
\end{equation}
with $\nabla = (\partial/\partial x, \partial/\partial y)$.  The
non-adiabatic couplings, $\mathbf{d}_{ij}^{(1)}$ and $d_{ij}^{(2)}$
are the vector and scalar derivative coupling matrix elements, and
$\hat W_{\pm}$ are the matrix elements of the adiabatic potential
$\mathbf{\hat W}=\mathbf{U\hat VU}^{\dagger}$:
\begin{equation}
  \label{eq:adiab-pots}
  \hat W_{\pm} = \frac{\hat V_{11} + \hat V_{22}}{2} \pm
  \frac{1}{2}\sqrt{(\hat V_{11} - \hat V_{22})^2 + 4\hat V_{12}^2}. 
\end{equation}

\begin{figure}
  \centering
  \includegraphics[width=0.5\textwidth]{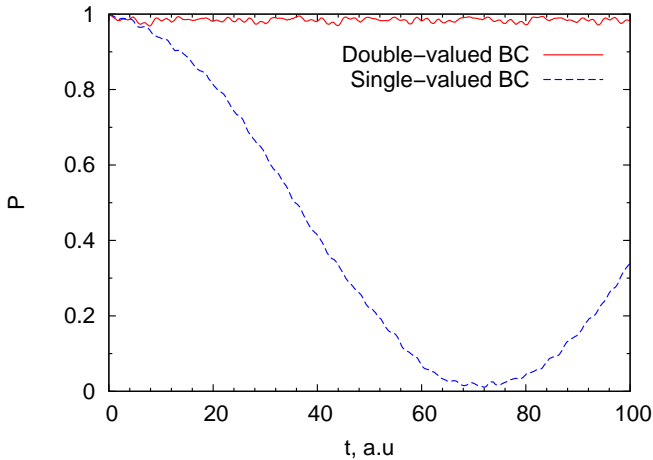}
  \caption{The solutions of Eq.~(\ref{eq:tdse-nuc}) with double- and
    single-valued boundary conditions. The initial wavepacket is a
    simple Gaussian located in the left well; $P$ is the fraction of
    the wavepacket located in the left well. The model parameters are $\omega=2, a=3, c=4$.}
  \label{fig:double_vs_single}
\end{figure}

Equation~(\ref{eq:tdse-nuc}) has to be solved by imposing
double-value boundary conditions for $\hat X_{\alpha\alpha'}$ due to
the presence of the nontrivial \gls{GP}.  Single-valued boundary
conditions lead to a completely different dynamics as shown in
Fig.~\ref{fig:double_vs_single}.

To avoid \gls{GP} complications in this model one can simulate
dynamics for the diabatic nuclear density
\begin{eqnarray}
  \label{eq:nuc-den-dia}
  \hat \rho_{\alpha\alpha'} = \Braket{\alpha |{\hat\rho}| \alpha'} = \ket{\chi_{\alpha}^\text{dia}}\bra{\chi_{\alpha'}^\text{dia}},  
\end{eqnarray}
by solving the corresponding Liouville equation
\begin{equation}
  \label{eq:tdse-nuc-dia}
  i\hbar\frac{\partial \hat{\boldsymbol\rho}}{\partial t} 
  = [\hat{\mathbf{H}},\hat{\boldsymbol\rho}].
\end{equation}
Of course, this diabatic path cannot be strictly followed for real
molecules where the diabatic representation is not rigorously defined.
But in this work we will use the diabatic representation to generate
the exact quantum dynamics for our model problem.

\section{Quantum-classical Liouville formalism}
\label{sec:quant-class-liouv}

In what follows we review the main steps of the
WA-QCL\cite{Kapral:1999/jcp/8919} and
AW-QCL~\cite{Horenko:2002/jcp/11075,Ando:2003/jcp/10399} derivations
to see the consequences of the double-valued character of the
adiabatic electronic states in both approaches. We consider only a
single set of nuclear coordinates associated with a particle of mass
$M$ to simplify our discussion and to remain close to that
in Sec.~\ref{sec:geometric-phase-two};
generalization to include additional nuclear coordinates is
straightforward.

\subsection{Wigner-then-adiabatic path}
\label{sec:from-full-electron}

In line with Ref.~\onlinecite{Kapral:1999/jcp/8919} we start with the
WT of the nuclear DOF in the Liouville equation for the total
density matrix in the coordinate representation
$\hat\rho(\bfr,\bfR,\bfr',\bfR') =
\Psi^{*}(\bfr,\bfR)\Psi(\bfr',\bfR')$,
\begin{equation}
  \label{eq:W_trans_Liouv}
  i\hbar\frac{\partial \hat\rho_W}{\partial t} = (\hat H \hat\rho)_W -
  (\hat\rho \hat H)_W. 
\end{equation}
Here, we used the Wigner representation for operators
\begin{equation}
  \label{eq:nuclear-Wigner}
  \hat A_W(\mathbf{R}_c, \mathbf{P}) = \int
  \left\langle\mathbf{R}_c + \frac{\mathbf{s}}{2}\right|\hat A
  \left|\mathbf{R}_c - \frac{\mathbf{s}}{2}\right\rangle
  e^{-\frac{i}{\hbar}\mathbf{s} \cdot \mathbf{P}} d\mathbf{s},
\end{equation}
where $\mathbf{R}_c = (\mathbf{R} + \mathbf{R}')/2$ is the position of
the ``coordinate centroid'', $\mathbf{s} = \mathbf{R} - \mathbf{R}'$,
and $\mathbf{P}$ is a parameter that can be associated with the classical momentum in the classical limit.  
Operator products $ \left(\hat A\hat B\right)_W$ can be
 transformed further using the Wigner-Moyal operator\cite{Imre:1967/jmp/1097}
$e^{-i\hbar\hat{\Lambda}/2}$ as
\begin{eqnarray}
  \label{eq:W_product}
  \left(\hat A\hat B\right)_W  & = & \hat A_W
  e^{-i\hbar\hat{\Lambda}/2} \hat B_W,
\end{eqnarray}
where $\hat{\Lambda}$ is the Poisson bracket operator $\hat{\Lambda} =
\overleftarrow\nabla_\mathbf{P} \cdot
\overrightarrow\nabla_{\mathbf{R}_c} -
\overleftarrow\nabla_{\mathbf{R}_c} \cdot
\overrightarrow\nabla_\mathbf{P}$, and the arrows indicate the 
directions in which the differential operators act. Using identity
(\ref{eq:W_product}) the Wigner-transformed Liouville equation becomes
\begin{equation}
  \label{eq:W_trans_Liouv_expan}
  i\hbar\frac{\partial \hat\rho_W}{\partial t} = \hat H_W e^{-i\hbar\hat{\Lambda}/2}
  \hat\rho_W - \hat\rho_W e^{-i\hbar\hat{\Lambda}/2} \hat H_W.
\end{equation}
The $\hat{\Lambda}$ operator acts only on the nuclear \gls{DOF}, thus,
to introduce a semi-classical description of nuclear dynamics we
expand the exponent in a Taylor series\footnote{A more physically appealing derivation 
proceeds by scaling the equation so that an expansion of the Liouville operator in small 
parameter $(m/M)^{1/2}$, where m and M are the characteristic masses of the quantum 
subsystem and bath DOF, can be carried out ( see Ref~\onlinecite{Kapral:1999/jcp/8919}).} 
and keep only terms linear in $\hbar$:
\begin{eqnarray}\notag
  i\hbar\frac{\partial \hat\rho_W}{\partial t} &=& \hat H_W \left(1-i\hbar\hat{\Lambda}/2\right) \hat\rho_W \\  \label{eq:W_trans_Liouv1}
  &&- \hat\rho_W \left(1-i\hbar\hat{\Lambda}/2\right) \hat H_W.  
\end{eqnarray}
The partially Wigner-transformed $H_W$ derived from \eq{eq:H_lvc} is quadratic
in $\mathbf{R}$ and $\mathbf{P}$, hence, the Wigner-transformed
Liouville equation~(\ref{eq:W_trans_Liouv1}) is exact for our model
case.

Both the Wigner-transformed density matrix $\hat\rho_W$ and
Hamiltonian $\hat H_W$ in Eq.~(\ref{eq:W_trans_Liouv1}) are still
quantum operators in the electronic subspace.  Let us project the
electronic \gls{DOF} on the adiabatic electronic eigenfuctions
$\ket{\psi_\alpha^\text{adi}(\mathbf{R}_c)}$
\begin{align}
  \label{eq:W_trans_Liouv_proj}
  i\hbar\frac{\partial \hat\rho_W^{\alpha\alpha'}}{\partial t} = {} &
  \bra{\psi_\alpha^\text{adi}}\hat H_W \left(1 -
    \frac{i\hbar}{2}\hat{\Lambda}\right)
  \hat\rho_W \ket{\psi_{\alpha'}^\text{adi}} \nonumber \\
  {} - {} & \bra{\psi_\alpha^\text{adi}} \hat\rho_W \left(1 -
    \frac{i\hbar}{2}\hat{\Lambda}\right) \hat
  H_W\ket{\psi_{\alpha'}^\text{adi}}.
\end{align}\sloppy
Inserting the resolution-of-the-identity operator in the electronic
subspace $ {\hat I} = \sum_\beta
\ket{\psi_\beta^\text{adi}(\mathbf{R}_c)}\bra{\psi_\beta^\text{adi}(\mathbf{R}_c)}$
between the operator products, we express
Eq.~(\ref{eq:W_trans_Liouv_proj}) in terms of operator matrices
defined as $\mathbf{A}_W \equiv \braket{\psi_\alpha^\text{adi}|\hat
  A_W |\psi_{\alpha'}^\text{adi}}$
\begin{align}
  \label{eq:QCL-matrix}
  \frac{\partial {\boldsymbol\rho}_W}{\partial t} & {} =
  -\frac{i}{\hbar}[\mathbf{H}_W, {\boldsymbol\rho}_W] -
  \frac{1}{2}\left(\mathbf{H}_W \hat{\Lambda} {\boldsymbol\rho}_W -
    {\boldsymbol\rho}_W \hat{\Lambda} \mathbf{H}_W\right) \nonumber \\
  & + \frac{1}{2}\Big([\mathbf{D}, \mathbf{H}_W] \cdot
  \nabla_\mathbf{P}{\boldsymbol\rho}_W +
  \nabla_\mathbf{P}{\boldsymbol\rho}_W \cdot [\mathbf{D},
  \mathbf{H}_W] \Big)
  \nonumber \\
  & -\frac{\mathbf{P}}{M}\cdot [\mathbf{D}, {\boldsymbol\rho}_W],
\end{align}
where $\mathbf{D}$ is the matrix of the vector derivative couplings
$\mathbf{d}_{\alpha\beta}^{(1)}$.  Equation (\ref{eq:QCL-matrix}) accounts for mutual interactions between the electronic and nuclear DOF. 
Note that all Wigner-transformed
operators depend on a single nuclear coordinate $\mathbf{R}_c$, and
even though the adiabatic functions
$\ket{\psi_\alpha^\text{adi}(\mathbf{R}_c)}$ are double-valued, the
diagonal density matrix elements $\braket{\psi_\alpha^\text{adi}|\hat
  \rho_W |\psi_{\alpha}^\text{adi}}$ are all single-valued functions of
$\mathbf{R}_c$. As for the off-diagonal elements
$\braket{\psi_\alpha^\text{adi}|\hat \rho_W |\psi_{\beta}^\text{adi}}$ (coherences),
in general, they are double-valued functions, because the \gls{GP}
related sign flip of the bra and ket components may not be
simultaneous for any closed contour in the $\mathbf{R}_c$ subspace. An
example of such situation is a three-state model where two electronic
states $\ket{\psi_{1,2}^\text{adi}}$ have a \gls{CI}, and the third
state $\ket{\psi_{3}^\text{adi}}$ has an avoided crossing with the
first two states.  In this example, the matrix elements
$\braket{\psi_\alpha^\text{adi}|\hat \rho_W
  |\psi_{3}^\text{adi}}_{\alpha=1,2}$ are double-valued functions
because the sign change in $\bra{\psi_{1,2}^\text{adi}}$ is not
compensated by that in $\ket{\psi_{3}^\text{adi}}$ on encircling the
\gls{CI}. Thus, generally, \eq{eq:QCL-matrix} must be simulated with
double-valued boundary conditions for the off-diagonal elements.
However, for problems where the parametric dependence of all adiabatic
states leads to the same \gls{GP} change after encircling any closed
contour all matrix elements in \eq{eq:QCL-matrix} are single-valued
functions, and \eq{eq:QCL-matrix} does not need double-valued boundary
conditions.  Our 2D \gls{LVC} model is an example where the latter
condition is satisfied because both functions $
\ket{\psi_1^\text{adi}(\mathbf{R}_c)}$ and
$\ket{\psi_2^\text{adi}(\mathbf{R}_c)}$ either change their sign or do
not depending on whether the \gls{CI} point is enclosed by the
contour.  At any case, all quantities in \eq{eq:QCL-matrix} are well
defined mathematically, and their physical interpretation is
well-known.\cite{Kapral:1999/jcp/8919}

\subsection{Adiabatic-then-Wigner path}
\label{sec:from-adiabatic}

We start by projecting the Liouville equation onto the adiabatic
electronic basis $\ket{\psi_\alpha^\text{adi}}$ to obtain
\begin{equation}
  \label{eq:LN-adi}
  i\hbar \frac{\partial {\hat X}_{\alpha\alpha'}}{\partial t} =
  \sum_{\beta} \left(\hat H^\text{adi}_{\alpha\beta}\,{\hat X}_{\beta\alpha'} 
    - {\hat X}_{\alpha\beta}\,\hat H^\text{adi}_{\beta\alpha'}\right),
\end{equation} 
where ${\hat H}^\text{adi}_{\alpha\beta}$ and ${\hat X}_{\alpha\beta}$
are multi-state generalizations of the two-state adiabatic
Hamiltonian~(\ref{eq:Hadi}) and nuclear
density~(\ref{eq:nuc-den-adi}).  Applying the WT to
Eq.~(\ref{eq:LN-adi}) we obtain
\begin{align}
  \label{eq:AW-QCL}
  i\hbar \frac{\partial \mathbf{X}_W}{\partial t} & {} =
  \mathbf{H}^\text{adi}_{W} e^{-i\hbar\hat{\Lambda}/2}
  \mathbf{X}_W -\mathbf{X}_W e^{-i\hbar\hat{\Lambda}/2}
  \mathbf{H}^\text{adi}_{W}.
\end{align}
For the double-valued adiabatic densities ${\hat X}_{\alpha\alpha'}$
the WT
\begin{equation}
  \label{eq:X-Wt}
  X_W^{\alpha\alpha'} = \int \Braket{\mathbf{R}_c
    +\frac{\mathbf{s}}{2} |\hat X_{\alpha\alpha'} | \mathbf{R}_c
    -\frac{\mathbf{s}}{2}} e^{-\frac{i}{\hbar}\mathbf{s} \cdot \mathbf{P}}d\SB 
\end{equation}
is not well defined, because, as is shown in the Appendix, this
operation is equivalent to the Fourier transform of a double-valued
function. Unless special care is taken at this step the
double-valued character of the ${\hat X}_{\alpha\alpha'}$ density will
be lost. Here, we follow previous derivations\cite{Horenko:2002/jcp/11075,
  Ando:2003/jcp/10399} of the AW-QCL equation,
noting that it is inconsistent with subtleties arising from a nontrivial
\gls{GP}.

The Wigner-transformed adiabatic Hamiltonian $\mathbf{H}_W^\text{adi}$
in \eq{eq:AW-QCL} differs from the $\mathbf{H}_W$ matrix only by the
WT of the non-adiabatic coupling operators
${\hat{\boldsymbol{\tau}}}_{\alpha\beta}$ [see
Eq.~(\ref{eq:NAC-def})],
\begin{equation}
  \label{eq:H_adi_mat-def}
  \mathbf{H}^\text{adi}_W = \mathbf{H}_W +
  {\boldsymbol\tau}_W, 
\end{equation}
where the ${\boldsymbol\tau}_W$ matrix has elements
\begin{equation}
  \label{eq:tau_W}
  {\boldsymbol\tau}_W^{\alpha\alpha'} = -i\hbar\frac{\mathbf{P}}{M} \cdot
  \mathbf{d}^{(1)}_{\alpha\alpha'} +
  \frac{\hbar^2}{2M}\left(\nabla_{\mathbf{R}_c} \cdot
    \mathbf{d}^{(1)}_{\alpha\alpha'} - d^{(2)}_{\alpha\alpha'}\right).
\end{equation}
The last term in the parentheses defines a positive-definite matrix
$\mathbf{K}$ with elements
\begin{equation}
  \label{eq:K-def}
  K_{\alpha\alpha'} =
  \frac{\hbar^2}{2M}\left(\nabla \cdot
    \mathbf{d}^{(1)}_{\alpha\alpha'} - d^{(2)}_{\alpha\alpha'}\right)
  = \frac{\hbar^2}{2M}\braket{\nabla\psi_{\alpha}^\text{adi} |
    \nabla\psi_{\alpha'}^\text{adi}}.
\end{equation}
The explicit form of the adiabatic Hamiltonian obtained from
Eqs.~(\ref{eq:H_adi_mat-def}--\ref{eq:K-def}) is
\begin{equation}
  \label{eq:H_adi_mat-res}
  \mathbf{H}^\text{adi}_W = \mathbf{H}_W  - i\hbar\frac{\mathbf{P}}{M}
  \cdot \mathbf{D} + \hbar^2\mathbf{K}.
\end{equation}
Expanding the Wigner-Moyal exponent in the Taylor series and keeping
only first order terms in $\hbar$ in Eq.~(\ref{eq:AW-QCL}) we obtain
the AW-QCL equation,
\begin{align}
  \notag \frac{\partial \mathbf{X}_W}{\partial t} & {} =
  -\frac{i}{\hbar}\left[\mathbf{H}_W , \mathbf{X}_W \right] -
  \frac{1}{2}\left(\mathbf{H}_W \hat{\Lambda} \mathbf{X}_W -
    \mathbf{X}_W \hat{\Lambda} \mathbf{H}_W \right)
  \\ \label{eq:AW-QCL-final} & -\frac{\mathbf{P}}{M}\cdot\left[
    \mathbf{D},\mathbf{X}_W\right].
\end{align}
It is important to note that \eq{eq:AW-QCL-final} is no longer an
exact evolution equation for the quantum dynamics of our 2D \gls{LVC}
model. This is in contrast to Eq.~(\ref{eq:W_trans_Liouv1}) which is
fully equivalent to the quantum Liouville equation for this model. 
The difference stems from two sources: First, the adiabatic Hamiltonian [\eq{eq:H_adi_mat-res}] 
is not a quadratic function of the nuclear coordinates, and consequently the truncation 
of the Wigner--Moyal exponent is no longer exact. Second, the AW-QCL derivation
ignores the double-valued character of the nuclear density, which results in 
\eq{eq:AW-QCL-final} that corresponds to the evolution of 
a nuclear density with the incorrect (single-valued) boundary conditions. 
Comparing Eqs.~(\ref{eq:QCL-matrix}) and (\ref{eq:AW-QCL-final}) reveals that
both sources of the difference between the AW-QCL and WA-QCL approaches 
can be related to the third term in \eq{eq:QCL-matrix}. 
Since this term accounts for direct changes in the nuclear momenta 
as a result of coupling to the electronic DOF, one might expect significant 
influence of this term on the nuclear dynamics.  Moreover, as we shall 
see below, neglecting this term leads to loss of GP related features in the nuclear 
dynamics.


\section{Numerical simulations}
\label{sec:num-sim}

\begin{figure}
  \centering
  \includegraphics[width=0.5\textwidth]{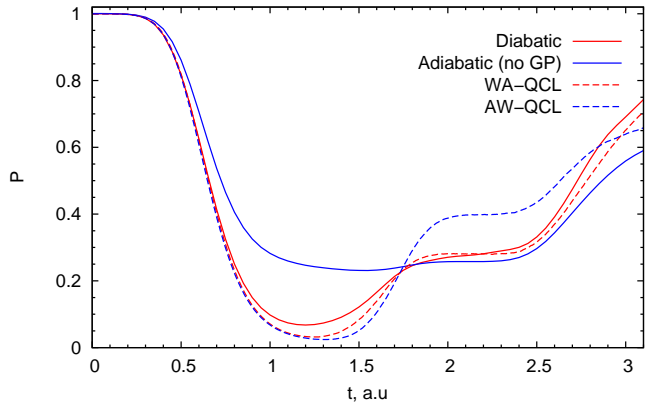}
  \caption{The fraction of the probability density remaining in the region $x < 0$ ($P$) 
  for the 2D \gls{LVC} model simulated with different methods.}
  \label{fig:podyn}
\end{figure}

\begin{figure}
  \centering
  \includegraphics[width=0.5\textwidth]{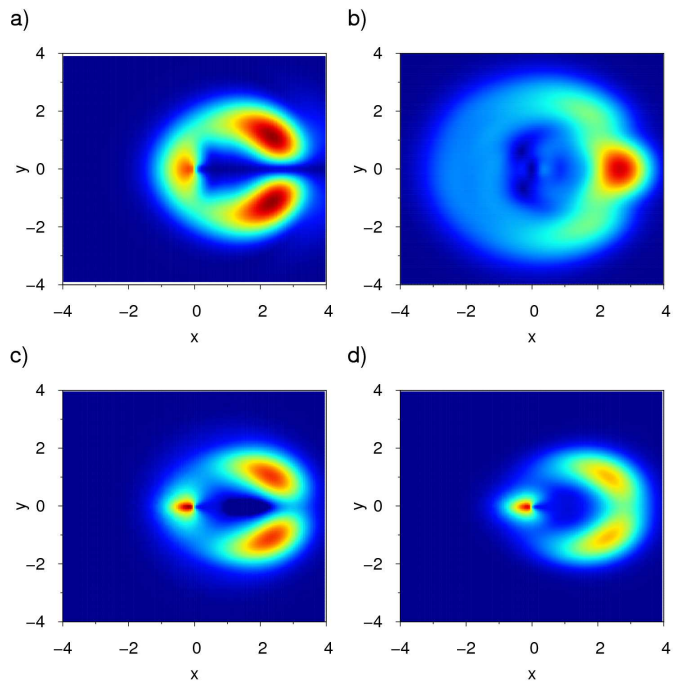}
  \caption{Snapshots of the square root of the probability density at
    t = 1.6 a.u.\ for the 2D \gls{LVC} model dynamics: a) fully
    quantum (diabatic) , b) quantum adiabatic (no \gls{GP}), c)
    WA-QCL, d) AW-QCL.}
  \label{fig:den}
\end{figure}

To quantify the difference between the WA-QCL and AW-QCL approaches
and to assess the relative importance of \gls{GP} effects we simulate non-adiabatic dynamics 
for the 2D \gls{LVC} model.  Our choice of 
initial conditions and the model parameters aims to minimize 
differences between the WA-QCL and AW-QCL approaches that are related to  
the omission of the higher order quantum contributions in AW-QCL. 
We employ the following 
set of parameters for the 2D \gls{LVC} model
[Eqs.~(\ref{eq:H_lvc}--\ref{eq:diab-me-22})]: $\omega = 2$, $a = 1$,
$c = 4$, $M=1$. The initial density is taken as a product of two Gaussian
wave packets centred at the minimum of the $\ket{1}$ diabatic surface
$(-a/2, 0)$ with widths $\sigma = \sqrt{2/\omega}$ and an initial
momentum $(5/2, 0)$ (motion towards the positive $x$ values).  
This setup results in a relatively high initial energy and reduction of quantum tunnelling. 
We consider four different approaches to non-adiabatic dynamics: a) quantum diabatic
[\eq{eq:tdse-nuc-dia}], b) quantum adiabatic without \gls{GP}
[\eq{eq:tdse-nuc}], c) WA-QCL, d) AW-QCL.  We investigate two
properties: the time evolution of the fraction of the nuclear
density remaining in the region $x <0$ (Fig.~\ref{fig:podyn}), and the 2D nuclear adiabatic
density at a given time (Fig.~\ref{fig:den}). The former is an integral property, which
illustrates differences in dynamics, while the latter zooms into these
differences at a given time.

As expected, the WA-QCL dynamics is very close to that of the full
quantum dynamics (see Fig.~\ref{fig:podyn}); the differences are due mainly to rather modest
convergence of the WA-QCL density with respect to the number of
classical trajectories in the employed Trotter-based simulation algorithm\cite{Kernan:2008ge} 
and use of the ``momentum jump'' approximation\cite{Kapral:1999/jcp/8919} for the last two terms in \eq{eq:QCL-matrix}.
Both approaches predict almost full transfer
of the density from the initial $x<0$ region to the $x>0$ region in a
time of $1.3$ a.u.\ with some degree of recurrence at later times. By
contrast, the quantum adiabatic dynamics exhibits less prominent
population transfer. Results of AW-QCL differ from those of the other
methods, which means that effects from the Wigner-Moyal exponent truncation  are
still comparable to \gls{GP} effects.  To elucidate the differences between
the methods we present the two-dimensional nuclear density at $t=1.6$
a.u. (Fig.~\ref{fig:den}).
The fully quantum model with \gls{GP} [Fig.~\ref{fig:den}(a)] and the 
WA-QCL model [Fig.~\ref{fig:den}(c)]  develop
a nodal line in the region $x >0$. This prominent feature is
completely absent in the quantum adiabatic dynamics
[Fig.~\ref{fig:den}(b)] and, thus, can be seen as a manifestation of
the \gls{GP}. \cite{Schon:1994/cpl/55, Ferretti:1996/jcp/5517,
  Kelly:2010/jcp/084502, Loic:2013un} Moreover, the adiabatic model
predicts completely spurious constructive interference between parts
of the wave packet that skirt the \gls{CI} point from opposite sides,
giving rise to a peak in the nuclear density at $x \approx 2.5$
a.u. [see Fig.~\ref{fig:den}(b)]. The AW-QCL density is somewhat
between those from both quantum models: it has a visible dip along the
$y=0$ line, however, this dip is not as pronounced as for the models
accounting for the \gls{GP}. The AW-QCL density profile suggests that
this approach does not produce any nuclear interference. To support
this assertion we considered different relations between two spatially
separated Gaussian wave packets.
\begin{figure}
  \centering
  \includegraphics[width=0.5\textwidth]{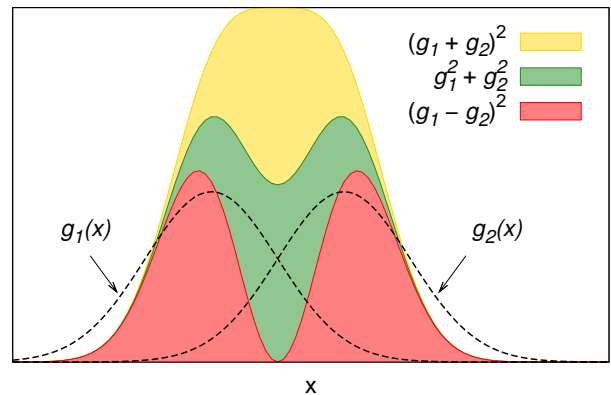}
  \caption{Total density of two Gaussian wave packets $g_1$ and $g_2$
    in cases of (gold) constructive interference, (red) destructive
    interference, (green) no interference.}
  \label{fig:interf}
\end{figure}
Figure~\ref{fig:interf} shows that in the absence of interference the
total density may have a central minimum, which, however, is not as
deep as in the case of destructive interference. This result also
indicates that electronic transitions constitute the dominant 
quantum effect in the AW-QCL dynamics, while quantum nuclear 
interference within each electronic state is negligible.

\section{Conclusions}
\label{sec:conclusions}

The analysis of the QCL dynamics for the 2D \gls{LVC} model presented above showed that
 the WA-QCL equation is the only approach where
complications arising from a non-trivial \gls{GP} in the adiabatic
representation do not make formalism ill-defined.  In situations when
the \gls{GP} behavior of the electronic adiabatic states involved in
the dynamics is similar (e.g., \gls{CI} of two electronic states), the
WA-QCL approach can treat \gls{GP} effects exactly without imposing
double-valued boundary conditions.  In contrast, the AW-QCL approach
involves the WT integral of the adiabatic
density operator with a double-valued kernel. Practically, this
transformation is equivalent to the Fourier transformation of a
double-valued function, which is not a well-defined procedure. Yet, if one proceeds 
with the Fourier transformation ignoring the double-valued character of the quantum 
density, the resulting AW-QCL equation describes the quantum density 
with the incorrect single-valued boundary conditions. Thus, the AW-QCL 
approach is plagued by the same problems as the quantum dynamics 
in the adiabatic representation without including the GP (see \fig{fig:double_vs_single}). 

Interestingly, the AW-QCL and WA-QCL equations differ only by one term. 
This term is responsible for differences associated with the Moyal--Wigner exponent 
truncation in different representations and GP effects.
Further separation of the GP-related terms from those corresponding to the Moyal--Wigner
exponent truncation does not seem to be feasible. 
However, this analysis suggests why the previous numerical assessment\cite{Ando:2003/jcp/10399} 
of the two approaches did not find a significant difference between results for the spin-boson model 
where \gls{GP} effects are absent.
 
It is worth noting that in contrast to the quantum adiabatic picture,
where the \gls{GP} appears in the form of non-trivial boundary
conditions, the WA-QCL approach confronts the problem of the \gls{GP} in much milder
fashion. The WT maps similar \gls{GP} behavior of different electronic states into the third term of the 
WA-QCL dynamical equation (\ref{eq:QCL-matrix}). Thus,  the only components 
that are left with the double-valued boundary conditions are coherences of electronic 
states that have different \gls{GP} behaviors. An interesting question for future
investigation is whether the trajectory based techniques used to simulate
\eq{eq:QCL-matrix} will be able to handle the double-valued character of the coherences 
in general multi-level systems owing to the space locality of a propagated object?

   It should be also emphasized that the WA-QCL approach, though accounting
   for the \gls{GP} in our model, cannot be reduced to the Mead and
   Truhlar~\cite{Mead:1979/jcp/2284} treatment of the \gls{GP}. This can be easily 
   seen by considering a single electronic state version of Eq.~(\ref{eq:QCL-matrix}) that 
   describes classical dynamics of nuclear DOF on a given electronic potential
   without any GP terms, which inevitably appear in the Mead and Truhlar treatment.

All quantities that are required to perform WA-QCL dynamics, such as
the adiabatic potentials and the derivative couplings, are readily
available from first-principles quantum chemistry calculations. Thus,
the WA-QCL formalism is probably the only formalism that can naturally account for \gls{GP} effects
 in two electronic state crossing problems with on-the-fly quantum chemistry calculations.

\begin{acknowledgments}
  RK and AFI acknowledge support from the Natural Science and
  Engineering Research Council (NSERC) of Canada through the Discovery Grants
  Program.
\end{acknowledgments}

\appendix

\section{Wigner transformation of adiabatic nuclear densities}
\label{app}

Here we consider the \gls{WT} of the adiabatic densities,
\begin{eqnarray}
  \notag
  X_W^{\alpha\alpha'}(\mathbf{P};\mathbf{R}_c) &=& \int \Braket{\mathbf{R}_c
    +\frac{\mathbf{s}}{2} |\hat X_{\alpha\alpha'} | \mathbf{R}_c
    -\frac{\mathbf{s}}{2}} e^{-\frac{i}{\hbar}\mathbf{s} \cdot \mathbf{P}}d\SB \\ \notag
  & = & \int
  \chi_\alpha\left(\mathbf{R}_c+\frac{\mathbf{s}}{2}\right)^{*}
  \chi_{\alpha'} \left(\mathbf{R}_c-\frac{\mathbf{s}}{2}\right) 
  e^{-\frac{i}{\hbar}\mathbf{s} \cdot \mathbf{P}}d\SB \\ 
  \label{eq:FTaa} 
  & = & \int
  f_{\alpha\alpha'}(\mathbf{s};\mathbf{R}_c)e^{-\frac{i}{\hbar}\mathbf{s} \cdot \mathbf{P}}d\SB. 
\end{eqnarray}
Equation~(\ref{eq:FTaa}) illustrates that the \gls{WT} can be
formulated as the Fourier transform (FT) of the $\mathbf{s}$ variable
for functions $f_{\alpha\alpha'}(\mathbf{s};\mathbf{R}_c)$.  In what
follows we will show that for the 2D \gls{LVC} model the
$f_{\alpha\alpha'}(\mathbf{s};\mathbf{R}_c)$ are double-valued
functions of the $\mathbf{s}$ variable, and therefore, the FT in
\eq{eq:FTaa} is not well defined.

\sloppy
Even for our model we do not have an exact analytical representation
of the $f_{\alpha\alpha'}(\mathbf{s};\mathbf{R}_c)$ functions, but we
know that the total density function ${ \hat\rho = \sum_{\alpha\alpha'}
\ket{\psi_{\alpha'}^{\rm adi}(\mathbf{R}_c-\mathbf{s}/2)}
f_{\alpha\alpha'}(\mathbf{s};\mathbf{R}_c) \bra{\psi_{\alpha}^{\rm
    adi}(\mathbf{R}_c+\mathbf{s}/2)} }$ is always a single-valued
function. Thus, if ${ f_{\alpha\alpha'}(\mathbf{s};\mathbf{R}_c) }$ are
double-valued then ${ \ket{\psi_{\alpha'}^{\rm
    adi}(\mathbf{R}_c-\mathbf{s}/2)}\bra{\psi_{\alpha}^{\rm
    adi}(\mathbf{R}_c+\mathbf{s}/2)} }$ must be too, and \emph{vice
  versa}.  Hence, we will show that the electronic components
$\ket{\psi_{\alpha'}^{\rm
    adi}(\mathbf{R}_c-\mathbf{s}/2)}\bra{\psi_{\alpha}^{\rm
    adi}(\mathbf{R}_c+\mathbf{s}/2)}$ are double-valued functions of
the $\mathbf{s}$ parameter. This is not a trivial check because the 
double-valued characters of the bra and ket components can potentially
compensate each other.  We will demonstrate by explicit construction 
that there exist at least some closed contours for the $\mathbf{s}$ parameter which, 
for a fixed $\mathbf{R}_c$, causes only $\ket{\psi_{\alpha'}^{\rm
    adi}(\mathbf{R}_c-\mathbf{s}/2)}$ or $\bra{\psi_{\alpha}^{\rm
    adi}(\mathbf{R}_c+\mathbf{s}/2)}$ to change its sign.

For the 2D \gls{LVC} model the functions $\ket{\psi_{\alpha'}^{\rm
    adi}(\mathbf{R}_c-\mathbf{s}/2)}$ and $\bra{\psi_{\alpha}^{\rm
    adi}(\mathbf{R}_c+\mathbf{s}/2)}$ are defined by
\eqs{eq:psi_el_ad_1} and (\ref{eq:psi_el_ad_2}). To construct desired
contours we fixed $\mathbf{R_c}=(x_c,y_c)$ and obtained the $(x,y)$
coordinates of the $\mathbf{R}_c-\mathbf{s}/2$ and
$\mathbf{R}_c+\mathbf{s}/2$ contours by taking the real and imaginary
parts of
\begin{eqnarray}\label{eq:cont}
z_{\pm s} = z_c \pm (d+re^{i\phi})/2.
\end{eqnarray}
\begin{figure}
  \centering
  \includegraphics[width=0.5\textwidth]{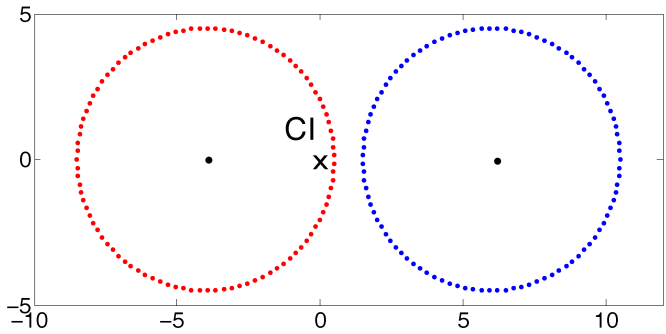}
  \caption{Two contours $z_{\pm s}$ of \eq{eq:cont}: the red contour
    ($z_{-s}$) encircles the \gls{CI} point, and the blue contour
    ($z_{+s}$) does not.  Parameters were set to $z_c = x_c+iy_c=1$,
    $d=10$, $r=9$.}
  \label{fig:circ2}
\end{figure}
Figure~\ref{fig:circ2} shows that this choice generates two contours
which are topologically different with respect to the \gls{CI}
point. Thus, the contour that encompasses the \gls{CI} produces
a change in the mixing angle $\theta$ [\eq{eq:theta}] by $2\pi$,
while moving along the other contour returns $\theta$ to its initial
value (see Fig.~\ref{fig:theta}).
\begin{figure}
  \centering
  \includegraphics[width=0.5\textwidth]{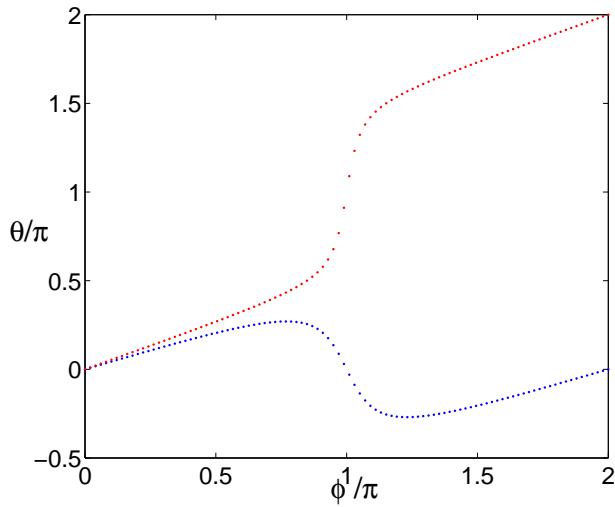}
  \caption{$\theta$-$\phi$ relation for the contours on
    \fig{fig:circ2}: the red (blue) dots follow the contour that
    encircles (does not encircle) the \gls{CI} point. Parameters of
    the Hamiltonian (\ref{eq:H_lvc}) are chosen so that $\theta={\rm
      arctan}(y/x)$}
  \label{fig:theta}
\end{figure}
Adiabatic transport along these contours will lead to a sign change
in all products of the electronic components $\ket{\psi_{\alpha'}^{\rm
    adi}(\mathbf{R}_c-\mathbf{s}/2)}\bra{\psi_{\alpha}^{\rm
    adi}(\mathbf{R}_c+\mathbf{s}/2)}$ and, thus, this concludes our
illustration of the double-valued character of the
$f_{\alpha\alpha'}(\mathbf{s};\mathbf{R}_c)$ functions.
 

\begin{thebibliography}{46}%
\makeatletter
\providecommand \@ifxundefined [1]{%
 \@ifx{#1\undefined}
}%
\providecommand \@ifnum [1]{%
 \ifnum #1\expandafter \@firstoftwo
 \else \expandafter \@secondoftwo
 \fi
}%
\providecommand \@ifx [1]{%
 \ifx #1\expandafter \@firstoftwo
 \else \expandafter \@secondoftwo
 \fi
}%
\providecommand \natexlab [1]{#1}%
\providecommand \enquote  [1]{``#1''}%
\providecommand \bibnamefont  [1]{#1}%
\providecommand \bibfnamefont [1]{#1}%
\providecommand \citenamefont [1]{#1}%
\providecommand \href@noop [0]{\@secondoftwo}%
\providecommand \href [0]{\begingroup \@sanitize@url \@href}%
\providecommand \@href[1]{\@@startlink{#1}\@@href}%
\providecommand \@@href[1]{\endgroup#1\@@endlink}%
\providecommand \@sanitize@url [0]{\catcode `\\12\catcode `\$12\catcode
  `\&12\catcode `\#12\catcode `\^12\catcode `\_12\catcode `\%12\relax}%
\providecommand \@@startlink[1]{}%
\providecommand \@@endlink[0]{}%
\providecommand \url  [0]{\begingroup\@sanitize@url \@url }%
\providecommand \@url [1]{\endgroup\@href {#1}{\urlprefix }}%
\providecommand \urlprefix  [0]{URL }%
\providecommand \Eprint [0]{\href }%
\providecommand \doibase [0]{http://dx.doi.org/}%
\providecommand \selectlanguage [0]{\@gobble}%
\providecommand \bibinfo  [0]{\@secondoftwo}%
\providecommand \bibfield  [0]{\@secondoftwo}%
\providecommand \translation [1]{[#1]}%
\providecommand \BibitemOpen [0]{}%
\providecommand \bibitemStop [0]{}%
\providecommand \bibitemNoStop [0]{.\EOS\space}%
\providecommand \EOS [0]{\spacefactor3000\relax}%
\providecommand \BibitemShut  [1]{\csname bibitem#1\endcsname}%
\let\auto@bib@innerbib\@empty
\bibitem [{\citenamefont {Migani}\ and\ \citenamefont
  {Olivucci}(2004)}]{Migani:2004/271}%
  \BibitemOpen
  \bibfield  {author} {\bibinfo {author} {\bibfnamefont {A.}~\bibnamefont
  {Migani}}\ and\ \bibinfo {author} {\bibfnamefont {M.}~\bibnamefont
  {Olivucci}},\ }in\ \href@noop {} {\emph {\bibinfo {booktitle} {{Conical
  Intersection Electronic Structure, Dynamics and Spectroscopy}}}},\ \bibinfo
  {editor} {edited by\ \bibinfo {editor} {\bibfnamefont {W.}~\bibnamefont
  {Domcke}}, \bibinfo {editor} {\bibfnamefont {D.~R.}\ \bibnamefont {Yarkony}},
  \ and\ \bibinfo {editor} {\bibfnamefont {H.}~\bibnamefont {K\"{o}ppel}}}\
  (\bibinfo  {publisher} {World Scientific},\ \bibinfo {address} {Singapore},\
  \bibinfo {year} {2004})\ p.\ \bibinfo {pages} {271}\BibitemShut {NoStop}%
\bibitem [{\citenamefont {Yarkony}(1996)}]{Yarkony:1996/rmp/985}%
  \BibitemOpen
  \bibfield  {author} {\bibinfo {author} {\bibfnamefont {D.~R.}\ \bibnamefont
  {Yarkony}},\ }\href {\doibase 10.1103/RevModPhys.68.985} {\bibfield
  {journal} {\bibinfo  {journal} {Rev. Mod. Phys.}\ }\textbf {\bibinfo {volume}
  {68}},\ \bibinfo {pages} {985} (\bibinfo {year} {1996})}\BibitemShut
  {NoStop}%
\bibitem [{\citenamefont {Blancafort}, \citenamefont {Hunt},\ and\
  \citenamefont {Robb}(2005)}]{Blancafort:2005/JACS/3391}%
  \BibitemOpen
  \bibfield  {author} {\bibinfo {author} {\bibfnamefont {L.}~\bibnamefont
  {Blancafort}}, \bibinfo {author} {\bibfnamefont {P.}~\bibnamefont {Hunt}}, \
  and\ \bibinfo {author} {\bibfnamefont {M.~A.}\ \bibnamefont {Robb}},\ }\href
  {\doibase 10.1021/ja043879h} {\bibfield  {journal} {\bibinfo  {journal} {J.
  Am. Chem. Soc.}\ }\textbf {\bibinfo {volume} {127}},\ \bibinfo {pages} {3391}
  (\bibinfo {year} {2005})}\BibitemShut {NoStop}%
\bibitem [{\citenamefont {Blancafort}\ \emph {et~al.}(2001)\citenamefont
  {Blancafort}, \citenamefont {Jolibois}, \citenamefont {Olivucci},\ and\
  \citenamefont {Robb}}]{Blancafort:2001/JACS/722}%
  \BibitemOpen
  \bibfield  {author} {\bibinfo {author} {\bibfnamefont {L.}~\bibnamefont
  {Blancafort}}, \bibinfo {author} {\bibfnamefont {F.}~\bibnamefont
  {Jolibois}}, \bibinfo {author} {\bibfnamefont {M.}~\bibnamefont {Olivucci}},
  \ and\ \bibinfo {author} {\bibfnamefont {M.~A.}\ \bibnamefont {Robb}},\
  }\href {\doibase 10.1021/ja003359w} {\bibfield  {journal} {\bibinfo
  {journal} {J. Am. Chem. Soc.}\ }\textbf {\bibinfo {volume} {123}},\ \bibinfo
  {pages} {722} (\bibinfo {year} {2001})}\BibitemShut {NoStop}%
\bibitem [{\citenamefont {Izmaylov}\ \emph {et~al.}(2011)\citenamefont
  {Izmaylov}, \citenamefont {{Mendive-Tapia}}, \citenamefont {Bearpark},
  \citenamefont {Robb}, \citenamefont {Tully},\ and\ \citenamefont
  {Frisch}}]{Izmaylov:2011/jcp/234106}%
  \BibitemOpen
  \bibfield  {author} {\bibinfo {author} {\bibfnamefont {A.~F.}\ \bibnamefont
  {Izmaylov}}, \bibinfo {author} {\bibfnamefont {D.}~\bibnamefont
  {{Mendive-Tapia}}}, \bibinfo {author} {\bibfnamefont {M.~J.}\ \bibnamefont
  {Bearpark}}, \bibinfo {author} {\bibfnamefont {M.~A.}\ \bibnamefont {Robb}},
  \bibinfo {author} {\bibfnamefont {J.~C.}\ \bibnamefont {Tully}}, \ and\
  \bibinfo {author} {\bibfnamefont {M.~J.}\ \bibnamefont {Frisch}},\ }\href
  {\doibase 10.1063/1.3667203} {\bibfield  {journal} {\bibinfo  {journal} {J.
  Chem. Phys.}\ }\textbf {\bibinfo {volume} {135}},\ \bibinfo {pages} {234106}
  (\bibinfo {year} {2011})}\BibitemShut {NoStop}%
\bibitem [{\citenamefont {Longuet-Higgins}\ \emph {et~al.}(1958)\citenamefont
  {Longuet-Higgins}, \citenamefont {Opik}, \citenamefont {Pryce},\ and\
  \citenamefont {Sack}}]{LonguetHigg:1958/rspa/1}%
  \BibitemOpen
  \bibfield  {author} {\bibinfo {author} {\bibfnamefont {H.~C.}\ \bibnamefont
  {Longuet-Higgins}}, \bibinfo {author} {\bibfnamefont {U.}~\bibnamefont
  {Opik}}, \bibinfo {author} {\bibfnamefont {M.~H.~L.}\ \bibnamefont {Pryce}},
  \ and\ \bibinfo {author} {\bibfnamefont {R.~A.}\ \bibnamefont {Sack}},\
  }\href {\doibase 10.1098/rspa.1958.0022} {\bibfield  {journal} {\bibinfo
  {journal} {Proc. R. Soc. A}\ }\textbf {\bibinfo {volume} {244}},\ \bibinfo
  {pages} {1} (\bibinfo {year} {1958})}\BibitemShut {NoStop}%
\bibitem [{\citenamefont {Berry}(1984)}]{Berry:1984/rspa/45}%
  \BibitemOpen
  \bibfield  {author} {\bibinfo {author} {\bibfnamefont {M.~V.}\ \bibnamefont
  {Berry}},\ }\href {\doibase 10.1098/rspa.1984.0023} {\bibfield  {journal}
  {\bibinfo  {journal} {Proc. R. Soc. A}\ }\textbf {\bibinfo {volume} {392}},\
  \bibinfo {pages} {45} (\bibinfo {year} {1984})}\BibitemShut {NoStop}%
\bibitem [{\citenamefont {Simon}(1983)}]{Simon:1983/prl/2167}%
  \BibitemOpen
  \bibfield  {author} {\bibinfo {author} {\bibfnamefont {B.}~\bibnamefont
  {Simon}},\ }\href {\doibase 10.1103/PhysRevLett.51.2167} {\bibfield
  {journal} {\bibinfo  {journal} {Phys. Rev. Lett.}\ }\textbf {\bibinfo
  {volume} {51}},\ \bibinfo {pages} {2167} (\bibinfo {year}
  {1983})}\BibitemShut {NoStop}%
\bibitem [{\citenamefont {Ham}(1987)}]{Ham:1987/prl/725}%
  \BibitemOpen
  \bibfield  {author} {\bibinfo {author} {\bibfnamefont {F.~S.}\ \bibnamefont
  {Ham}},\ }\href {\doibase 10.1103/PhysRevLett.58.725} {\bibfield  {journal}
  {\bibinfo  {journal} {Phys. Rev. Lett.}\ }\textbf {\bibinfo {volume} {58}},\
  \bibinfo {pages} {725} (\bibinfo {year} {1987})}\BibitemShut {NoStop}%
\bibitem [{\citenamefont {Berry}(1987)}]{Berry:1987/rspa/31}%
  \BibitemOpen
  \bibfield  {author} {\bibinfo {author} {\bibfnamefont {M.~V.}\ \bibnamefont
  {Berry}},\ }\href {\doibase 10.1098/rspa.1987.0131} {\bibfield  {journal}
  {\bibinfo  {journal} {Proc. R. Soc. A}\ }\textbf {\bibinfo {volume} {414}},\
  \bibinfo {pages} {31} (\bibinfo {year} {1987})}\BibitemShut {NoStop}%
\bibitem [{\citenamefont {Aitchison}(1988)}]{Aitchison:1988/ps/12}%
  \BibitemOpen
  \bibfield  {author} {\bibinfo {author} {\bibfnamefont {I.~J.~R.}\
  \bibnamefont {Aitchison}},\ }\href {\doibase 10.1088/0031-8949/1988/T23/002}
  {\bibfield  {journal} {\bibinfo  {journal} {Phys. Scr.}\ }\textbf {\bibinfo
  {volume} {1988}},\ \bibinfo {pages} {12} (\bibinfo {year}
  {1988})}\BibitemShut {NoStop}%
\bibitem [{\citenamefont {Sch{\"{o}}n}\ and\ \citenamefont
  {K{\"{o}}ppel}(1994)}]{Schon:1994/cpl/55}%
  \BibitemOpen
  \bibfield  {author} {\bibinfo {author} {\bibfnamefont {J.}~\bibnamefont
  {Sch{\"{o}}n}}\ and\ \bibinfo {author} {\bibfnamefont {H.}~\bibnamefont
  {K{\"{o}}ppel}},\ }\href {\doibase 10.1016/0009-2614(94)01219-9} {\bibfield
  {journal} {\bibinfo  {journal} {Chem. Phys. Lett.}\ }\textbf {\bibinfo
  {volume} {231}},\ \bibinfo {pages} {55} (\bibinfo {year} {1994})}\BibitemShut
  {NoStop}%
\bibitem [{\citenamefont {Sch\"{o}n}\ and\ \citenamefont
  {K\"{o}ppel}(1995)}]{Schon:1995/jcp/9292}%
  \BibitemOpen
  \bibfield  {author} {\bibinfo {author} {\bibfnamefont {J.}~\bibnamefont
  {Sch\"{o}n}}\ and\ \bibinfo {author} {\bibfnamefont {H.}~\bibnamefont
  {K\"{o}ppel}},\ }\href {\doibase 10.1063/1.469988} {\bibfield  {journal}
  {\bibinfo  {journal} {J. Chem. Phys.}\ }\textbf {\bibinfo {volume} {103}},\
  \bibinfo {pages} {9292} (\bibinfo {year} {1995})}\BibitemShut {NoStop}%
\bibitem [{\citenamefont {Ryabinkin}\ and\ \citenamefont
  {Izmaylov}(2013)}]{Ryabinkin:2013/prl/220406}%
  \BibitemOpen
  \bibfield  {author} {\bibinfo {author} {\bibfnamefont {I.~G.}\ \bibnamefont
  {Ryabinkin}}\ and\ \bibinfo {author} {\bibfnamefont {A.~F.}\ \bibnamefont
  {Izmaylov}},\ }\href {\doibase 10.1103/PhysRevLett.111.220406} {\bibfield
  {journal} {\bibinfo  {journal} {Phys. Rev. Lett.}\ }\textbf {\bibinfo
  {volume} {111}},\ \bibinfo {pages} {220406} (\bibinfo {year}
  {2013})}\BibitemShut {NoStop}%
\bibitem [{\citenamefont {Joubert-Doriol}, \citenamefont {Ryabinkin},\ and\
  \citenamefont {Izmaylov}(2013)}]{Loic:2013un}%
  \BibitemOpen
  \bibfield  {author} {\bibinfo {author} {\bibfnamefont {L.}~\bibnamefont
  {Joubert-Doriol}}, \bibinfo {author} {\bibfnamefont {I.~G.}\ \bibnamefont
  {Ryabinkin}}, \ and\ \bibinfo {author} {\bibfnamefont {A.~F.}\ \bibnamefont
  {Izmaylov}},\ }\href@noop {} {\bibfield  {journal} {\bibinfo  {journal} {J.
  Chem. Phys.}\ } (\bibinfo {year} {2013})},\ \bibinfo {note} {in press},\
  \Eprint {http://arxiv.org/abs/arXiv:1310.2929} {arXiv:1310.2929} \BibitemShut
  {NoStop}%
\bibitem [{\citenamefont {Cederbaum}(2004)}]{Cederbaum:2004/CI}%
  \BibitemOpen
  \bibfield  {author} {\bibinfo {author} {\bibfnamefont {L.~S.}\ \bibnamefont
  {Cederbaum}},\ }in\ \href@noop {} {\emph {\bibinfo {booktitle} {{Conical
  Intersections}}}},\ \bibinfo {editor} {edited by\ \bibinfo {editor}
  {\bibfnamefont {W.}~\bibnamefont {Domcke}}, \bibinfo {editor} {\bibfnamefont
  {D.~R.}\ \bibnamefont {Yarkony}}, \ and\ \bibinfo {editor} {\bibfnamefont
  {H.}~\bibnamefont {K{\"o}ppel}}}\ (\bibinfo  {publisher} {World Scientific},\
  \bibinfo {address} {Singapore},\ \bibinfo {year} {2004})\ pp.\ \bibinfo
  {pages} {3--40}\BibitemShut {NoStop}%
\bibitem [{\citenamefont {Baer}(1975)}]{Baer:1975/cpl/112}%
  \BibitemOpen
  \bibfield  {author} {\bibinfo {author} {\bibfnamefont {M.}~\bibnamefont
  {Baer}},\ }\href {\doibase 10.1016/0009-2614(75)85599-0} {\bibfield
  {journal} {\bibinfo  {journal} {Chem. Phys. Lett.}\ }\textbf {\bibinfo
  {volume} {35}},\ \bibinfo {pages} {112} (\bibinfo {year} {1975})}\BibitemShut
  {NoStop}%
\bibitem [{\citenamefont {K{\"{o}}ppel}\ and\ \citenamefont
  {Schubert}(2006)}]{Koppel:2006/mp/1069}%
  \BibitemOpen
  \bibfield  {author} {\bibinfo {author} {\bibfnamefont {H.}~\bibnamefont
  {K{\"{o}}ppel}}\ and\ \bibinfo {author} {\bibfnamefont {B.}~\bibnamefont
  {Schubert}},\ }\href {\doibase 10.1080/00268970500417937} {\bibfield
  {journal} {\bibinfo  {journal} {Mol. Phys.}\ }\textbf {\bibinfo {volume}
  {104}},\ \bibinfo {pages} {1069} (\bibinfo {year} {2006})}\BibitemShut
  {NoStop}%
\bibitem [{\citenamefont {Van~Voorhis}\ \emph {et~al.}(2010)\citenamefont
  {Van~Voorhis}, \citenamefont {Kowalczyk}, \citenamefont {Kaduk},
  \citenamefont {Wang}, \citenamefont {Cheng},\ and\ \citenamefont
  {Wu}}]{Voorhis:2010/arpc/149}%
  \BibitemOpen
  \bibfield  {author} {\bibinfo {author} {\bibfnamefont {T.}~\bibnamefont
  {Van~Voorhis}}, \bibinfo {author} {\bibfnamefont {T.}~\bibnamefont
  {Kowalczyk}}, \bibinfo {author} {\bibfnamefont {B.}~\bibnamefont {Kaduk}},
  \bibinfo {author} {\bibfnamefont {L.-P.}\ \bibnamefont {Wang}}, \bibinfo
  {author} {\bibfnamefont {C.-L.}\ \bibnamefont {Cheng}}, \ and\ \bibinfo
  {author} {\bibfnamefont {Q.}~\bibnamefont {Wu}},\ }\href {\doibase
  10.1146/annurev.physchem.012809.103324} {\bibfield  {journal} {\bibinfo
  {journal} {Ann. Rev. Phys. Chem.}\ }\textbf {\bibinfo {volume} {61}},\
  \bibinfo {pages} {149} (\bibinfo {year} {2010})}\BibitemShut {NoStop}%
\bibitem [{\citenamefont {Sirjoosingh}\ and\ \citenamefont
  {Hammes-Schiffer}(2011)}]{Sirjoosingh:2011/jctc/2831}%
  \BibitemOpen
  \bibfield  {author} {\bibinfo {author} {\bibfnamefont {A.}~\bibnamefont
  {Sirjoosingh}}\ and\ \bibinfo {author} {\bibfnamefont {S.}~\bibnamefont
  {Hammes-Schiffer}},\ }\href {\doibase 10.1021/ct200356b} {\bibfield
  {journal} {\bibinfo  {journal} {J. Chem. Theory Comput.}\ }\textbf {\bibinfo
  {volume} {7}},\ \bibinfo {pages} {2831} (\bibinfo {year} {2011})}\BibitemShut
  {NoStop}%
\bibitem [{\citenamefont {Yarkony}(2012)}]{Yarkony:2012/cr/481}%
  \BibitemOpen
  \bibfield  {author} {\bibinfo {author} {\bibfnamefont {D.~R.}\ \bibnamefont
  {Yarkony}},\ }\href {\doibase 10.1021/cr2001299} {\bibfield  {journal}
  {\bibinfo  {journal} {Chem. Rev.}\ }\textbf {\bibinfo {volume} {112}},\
  \bibinfo {pages} {481} (\bibinfo {year} {2012})}\BibitemShut {NoStop}%
\bibitem [{\citenamefont {Subotnik}\ \emph {et~al.}(2009)\citenamefont
  {Subotnik}, \citenamefont {Cave}, \citenamefont {Steele},\ and\ \citenamefont
  {Shenvi}}]{Subotnik:2009/jcp/234102}%
  \BibitemOpen
  \bibfield  {author} {\bibinfo {author} {\bibfnamefont {J.~E.}\ \bibnamefont
  {Subotnik}}, \bibinfo {author} {\bibfnamefont {R.~J.}\ \bibnamefont {Cave}},
  \bibinfo {author} {\bibfnamefont {R.~P.}\ \bibnamefont {Steele}}, \ and\
  \bibinfo {author} {\bibfnamefont {N.}~\bibnamefont {Shenvi}},\ }\href
  {\doibase 10.1063/1.3148777} {\bibfield  {journal} {\bibinfo  {journal} {J.
  Chem. Phys.}\ }\textbf {\bibinfo {volume} {130}},\ \bibinfo {pages} {234102}
  (\bibinfo {year} {2009})}\BibitemShut {NoStop}%
\bibitem [{\citenamefont {Opalka}\ and\ \citenamefont
  {Domcke}(2010)}]{Opalka:2010/jcp/154108}%
  \BibitemOpen
  \bibfield  {author} {\bibinfo {author} {\bibfnamefont {D.}~\bibnamefont
  {Opalka}}\ and\ \bibinfo {author} {\bibfnamefont {W.}~\bibnamefont
  {Domcke}},\ }\href {\doibase 10.1063/1.3382912} {\bibfield  {journal}
  {\bibinfo  {journal} {J. Chem. Phys.}\ }\textbf {\bibinfo {volume} {132}},\
  \bibinfo {pages} {154108} (\bibinfo {year} {2010})}\BibitemShut {NoStop}%
\bibitem [{\citenamefont {Troisi}\ and\ \citenamefont
  {Orlandi}(2003)}]{Troisi:2003/jcp/5356}%
  \BibitemOpen
  \bibfield  {author} {\bibinfo {author} {\bibfnamefont {A.}~\bibnamefont
  {Troisi}}\ and\ \bibinfo {author} {\bibfnamefont {G.}~\bibnamefont
  {Orlandi}},\ }\href {\doibase 10.1063/1.1555118} {\bibfield  {journal}
  {\bibinfo  {journal} {J. Chem. Phys.}\ }\textbf {\bibinfo {volume} {118}},\
  \bibinfo {pages} {5356} (\bibinfo {year} {2003})}\BibitemShut {NoStop}%
\bibitem [{\citenamefont {Papas}, \citenamefont {Schuurman},\ and\
  \citenamefont {Yarkony}(2008)}]{Papas:2008/jcp/124104}%
  \BibitemOpen
  \bibfield  {author} {\bibinfo {author} {\bibfnamefont {B.~N.}\ \bibnamefont
  {Papas}}, \bibinfo {author} {\bibfnamefont {M.~S.}\ \bibnamefont
  {Schuurman}}, \ and\ \bibinfo {author} {\bibfnamefont {D.~R.}\ \bibnamefont
  {Yarkony}},\ }\href {\doibase 10.1063/1.2978389} {\bibfield  {journal}
  {\bibinfo  {journal} {J. Chem. Phys.}\ }\textbf {\bibinfo {volume} {129}},\
  \bibinfo {pages} {124104} (\bibinfo {year} {2008})}\BibitemShut {NoStop}%
\bibitem [{\citenamefont {Nakamura}\ and\ \citenamefont
  {Truhlar}(2001)}]{Nakamura:2001/jcp/10353}%
  \BibitemOpen
  \bibfield  {author} {\bibinfo {author} {\bibfnamefont {H.}~\bibnamefont
  {Nakamura}}\ and\ \bibinfo {author} {\bibfnamefont {D.~G.}\ \bibnamefont
  {Truhlar}},\ }\href {\doibase 10.1063/1.1412879} {\bibfield  {journal}
  {\bibinfo  {journal} {J. Chem. Phys.}\ }\textbf {\bibinfo {volume} {115}},\
  \bibinfo {pages} {10353} (\bibinfo {year} {2001})}\BibitemShut {NoStop}%
\bibitem [{\citenamefont {Mead}\ and\ \citenamefont
  {Truhlar}(1979)}]{Mead:1979/jcp/2284}%
  \BibitemOpen
  \bibfield  {author} {\bibinfo {author} {\bibfnamefont {C.~A.}\ \bibnamefont
  {Mead}}\ and\ \bibinfo {author} {\bibfnamefont {D.~G.}\ \bibnamefont
  {Truhlar}},\ }\href {\doibase 10.1063/1.437734} {\bibfield  {journal}
  {\bibinfo  {journal} {J. Chem. Phys.}\ }\textbf {\bibinfo {volume} {70}},\
  \bibinfo {pages} {2284} (\bibinfo {year} {1979})}\BibitemShut {NoStop}%
\bibitem [{\citenamefont {Kendrick}\ and\ \citenamefont
  {Mead}(1995)}]{Kendrick:1995/jcp/4160}%
  \BibitemOpen
  \bibfield  {author} {\bibinfo {author} {\bibfnamefont {B.}~\bibnamefont
  {Kendrick}}\ and\ \bibinfo {author} {\bibfnamefont {C.~A.}\ \bibnamefont
  {Mead}},\ }\href {\doibase 10.1063/1.468544} {\bibfield  {journal} {\bibinfo
  {journal} {J. Chem. Phys.}\ }\textbf {\bibinfo {volume} {102}},\ \bibinfo
  {pages} {4160} (\bibinfo {year} {1995})}\BibitemShut {NoStop}%
\bibitem [{\citenamefont {Kendrick}, \citenamefont {Alden~Mead},\ and\
  \citenamefont {Truhlar}(2002)}]{Kendrick:2002/cp/31}%
  \BibitemOpen
  \bibfield  {author} {\bibinfo {author} {\bibfnamefont {B.~K.}\ \bibnamefont
  {Kendrick}}, \bibinfo {author} {\bibfnamefont {C.}~\bibnamefont
  {Alden~Mead}}, \ and\ \bibinfo {author} {\bibfnamefont {D.~G.}\ \bibnamefont
  {Truhlar}},\ }\href {\doibase 10.1016/S0301-0104(02)00281-1} {\bibfield
  {journal} {\bibinfo  {journal} {Chem. Phys.}\ }\textbf {\bibinfo {volume}
  {277}},\ \bibinfo {pages} {31} (\bibinfo {year} {2002})}\BibitemShut
  {NoStop}%
\bibitem [{\citenamefont {Kendrick}(2003{\natexlab{a}})}]{Kendrick:2003}%
  \BibitemOpen
  \bibfield  {author} {\bibinfo {author} {\bibfnamefont {B.~K.}\ \bibnamefont
  {Kendrick}},\ }in\ \href {\doibase 10.1142/9789812565464_0012} {\emph
  {\bibinfo {booktitle} {{Conical Intersections. Electronic Structure, Dynamics
  and Spectroscopy}}}},\ \bibinfo {series} {Advanced Series in Physical
  Chemistry}, Vol.~\bibinfo {volume} {15},\ \bibinfo {editor} {edited by\
  \bibinfo {editor} {\bibfnamefont {W.}~\bibnamefont {Domcke}}, \bibinfo
  {editor} {\bibfnamefont {D.~R.}\ \bibnamefont {Yarkony}}, \ and\ \bibinfo
  {editor} {\bibfnamefont {H.}~\bibnamefont {K\"oppel}}}\ (\bibinfo
  {publisher} {World Scientific},\ \bibinfo {address} {Singapore},\ \bibinfo
  {year} {2003})\ Chap.~\bibinfo {chapter} {12}, pp.\ \bibinfo {pages}
  {521--553}\BibitemShut {NoStop}%
\bibitem [{\citenamefont {Kendrick}\ and\ \citenamefont
  {Pack}(1996{\natexlab{a}})}]{Kendrick:1996/jcp/7475}%
  \BibitemOpen
  \bibfield  {author} {\bibinfo {author} {\bibfnamefont {B.}~\bibnamefont
  {Kendrick}}\ and\ \bibinfo {author} {\bibfnamefont {R.~T.}\ \bibnamefont
  {Pack}},\ }\href {\doibase 10.1063/1.471460} {\bibfield  {journal} {\bibinfo
  {journal} {J. Chem. Phys.}\ }\textbf {\bibinfo {volume} {104}},\ \bibinfo
  {pages} {7475} (\bibinfo {year} {1996}{\natexlab{a}})}\BibitemShut {NoStop}%
\bibitem [{\citenamefont {Kendrick}\ and\ \citenamefont
  {Pack}(1996{\natexlab{b}})}]{Kendrick:1996/jcp/7502}%
  \BibitemOpen
  \bibfield  {author} {\bibinfo {author} {\bibfnamefont {B.}~\bibnamefont
  {Kendrick}}\ and\ \bibinfo {author} {\bibfnamefont {R.~T.}\ \bibnamefont
  {Pack}},\ }\href {\doibase 10.1063/1.471461} {\bibfield  {journal} {\bibinfo
  {journal} {J. Chem. Phys.}\ }\textbf {\bibinfo {volume} {104}},\ \bibinfo
  {pages} {7502} (\bibinfo {year} {1996}{\natexlab{b}})}\BibitemShut {NoStop}%
\bibitem [{\citenamefont {Kendrick}(1997)}]{Kendrick:1997/prl/2431}%
  \BibitemOpen
  \bibfield  {author} {\bibinfo {author} {\bibfnamefont {B.}~\bibnamefont
  {Kendrick}},\ }\href {\doibase 10.1103/PhysRevLett.79.2431} {\bibfield
  {journal} {\bibinfo  {journal} {Phys. Rev. Lett.}\ }\textbf {\bibinfo
  {volume} {79}},\ \bibinfo {pages} {2431} (\bibinfo {year}
  {1997})}\BibitemShut {NoStop}%
\bibitem [{\citenamefont
  {Kendrick}(2003{\natexlab{b}})}]{Kendrick:2003/jpca/6739}%
  \BibitemOpen
  \bibfield  {author} {\bibinfo {author} {\bibfnamefont {B.~K.}\ \bibnamefont
  {Kendrick}},\ }\href {\doibase 10.1021/jp021865x} {\bibfield  {journal}
  {\bibinfo  {journal} {J. Phys. Chem. A}\ }\textbf {\bibinfo {volume} {107}},\
  \bibinfo {pages} {6739} (\bibinfo {year} {2003}{\natexlab{b}})}\BibitemShut
  {NoStop}%
\bibitem [{\citenamefont {Baer}(1997)}]{Baer:1997/jcp/2694}%
  \BibitemOpen
  \bibfield  {author} {\bibinfo {author} {\bibfnamefont {M.}~\bibnamefont
  {Baer}},\ }\href {\doibase 10.1063/1.474623} {\bibfield  {journal} {\bibinfo
  {journal} {J. Chem. Phys.}\ }\textbf {\bibinfo {volume} {107}},\ \bibinfo
  {pages} {2694} (\bibinfo {year} {1997})}\BibitemShut {NoStop}%
\bibitem [{\citenamefont {Xu}, \citenamefont {Baer},\ and\ \citenamefont
  {Varandas}(2000)}]{Xu:2000/jcp/2746}%
  \BibitemOpen
  \bibfield  {author} {\bibinfo {author} {\bibfnamefont {Z.}~\bibnamefont
  {Xu}}, \bibinfo {author} {\bibfnamefont {M.}~\bibnamefont {Baer}}, \ and\
  \bibinfo {author} {\bibfnamefont {A.~J.~C.}\ \bibnamefont {Varandas}},\
  }\href {\doibase 10.1063/1.480848} {\bibfield  {journal} {\bibinfo  {journal}
  {J. Chem. Phys.}\ }\textbf {\bibinfo {volume} {112}},\ \bibinfo {pages}
  {2746} (\bibinfo {year} {2000})}\BibitemShut {NoStop}%
\bibitem [{\citenamefont {Kendrick}, \citenamefont {Mead},\ and\ \citenamefont
  {Truhlar}(1999)}]{Kendrick:1999/jcp/7594}%
  \BibitemOpen
  \bibfield  {author} {\bibinfo {author} {\bibfnamefont {B.~K.}\ \bibnamefont
  {Kendrick}}, \bibinfo {author} {\bibfnamefont {C.~A.}\ \bibnamefont {Mead}},
  \ and\ \bibinfo {author} {\bibfnamefont {D.~G.}\ \bibnamefont {Truhlar}},\
  }\href {\doibase 10.1063/1.478670} {\bibfield  {journal} {\bibinfo  {journal}
  {J. Chem. Phys.}\ }\textbf {\bibinfo {volume} {110}},\ \bibinfo {pages}
  {7594} (\bibinfo {year} {1999})}\BibitemShut {NoStop}%
\bibitem [{\citenamefont {Kapral}(2006)}]{Kapral:2006kw}%
  \BibitemOpen
  \bibfield  {author} {\bibinfo {author} {\bibfnamefont {R.}~\bibnamefont
  {Kapral}},\ }\href@noop {} {\bibfield  {journal} {\bibinfo  {journal} {Annual
  Review of Physical Chemistry}\ }\textbf {\bibinfo {volume} {57}},\ \bibinfo
  {pages} {129} (\bibinfo {year} {2006})}\BibitemShut {NoStop}%
\bibitem [{\citenamefont {Kelly}\ and\ \citenamefont
  {Kapral}(2010)}]{Kelly:2010/jcp/084502}%
  \BibitemOpen
  \bibfield  {author} {\bibinfo {author} {\bibfnamefont {A.}~\bibnamefont
  {Kelly}}\ and\ \bibinfo {author} {\bibfnamefont {R.}~\bibnamefont {Kapral}},\
  }\href {\doibase 10.1063/1.3475773} {\bibfield  {journal} {\bibinfo
  {journal} {J. Chem. Phys.}\ }\textbf {\bibinfo {volume} {133}},\ \bibinfo
  {pages} {084502} (\bibinfo {year} {2010})}\BibitemShut {NoStop}%
\bibitem [{\citenamefont {Kapral}\ and\ \citenamefont
  {Ciccotti}(1999)}]{Kapral:1999/jcp/8919}%
  \BibitemOpen
  \bibfield  {author} {\bibinfo {author} {\bibfnamefont {R.}~\bibnamefont
  {Kapral}}\ and\ \bibinfo {author} {\bibfnamefont {G.}~\bibnamefont
  {Ciccotti}},\ }\href {\doibase 10.1063/1.478811} {\bibfield  {journal}
  {\bibinfo  {journal} {J. Chem. Phys.}\ }\textbf {\bibinfo {volume} {110}},\
  \bibinfo {pages} {8919} (\bibinfo {year} {1999})}\BibitemShut {NoStop}%
\bibitem [{\citenamefont {Horenko}\ \emph {et~al.}(2002)\citenamefont
  {Horenko}, \citenamefont {Salzmann}, \citenamefont {Schmidt},\ and\
  \citenamefont {Sch\"{u}tte}}]{Horenko:2002/jcp/11075}%
  \BibitemOpen
  \bibfield  {author} {\bibinfo {author} {\bibfnamefont {I.}~\bibnamefont
  {Horenko}}, \bibinfo {author} {\bibfnamefont {C.}~\bibnamefont {Salzmann}},
  \bibinfo {author} {\bibfnamefont {B.}~\bibnamefont {Schmidt}}, \ and\
  \bibinfo {author} {\bibfnamefont {C.}~\bibnamefont {Sch\"{u}tte}},\ }\href
  {\doibase 10.1063/1.1522712} {\bibfield  {journal} {\bibinfo  {journal} {J.
  Chem. Phys.}\ }\textbf {\bibinfo {volume} {117}},\ \bibinfo {pages} {11075}
  (\bibinfo {year} {2002})}\BibitemShut {NoStop}%
\bibitem [{\citenamefont {Ando}\ and\ \citenamefont
  {Santer}(2003)}]{Ando:2003/jcp/10399}%
  \BibitemOpen
  \bibfield  {author} {\bibinfo {author} {\bibfnamefont {K.}~\bibnamefont
  {Ando}}\ and\ \bibinfo {author} {\bibfnamefont {M.}~\bibnamefont {Santer}},\
  }\href {\doibase 10.1063/1.1574015} {\bibfield  {journal} {\bibinfo
  {journal} {J. Chem. Phys.}\ }\textbf {\bibinfo {volume} {118}},\ \bibinfo
  {pages} {10399} (\bibinfo {year} {2003})}\BibitemShut {NoStop}%
\bibitem [{\citenamefont {{\.I}mre}\ \emph {et~al.}(1967)\citenamefont
  {{\.I}mre}, \citenamefont {{\"{O}}zizmir}, \citenamefont {Rosenbaum},\ and\
  \citenamefont {Zweifel}}]{Imre:1967/jmp/1097}%
  \BibitemOpen
  \bibfield  {author} {\bibinfo {author} {\bibfnamefont {K.}~\bibnamefont
  {{\.I}mre}}, \bibinfo {author} {\bibfnamefont {E.}~\bibnamefont
  {{\"{O}}zizmir}}, \bibinfo {author} {\bibfnamefont {M.}~\bibnamefont
  {Rosenbaum}}, \ and\ \bibinfo {author} {\bibfnamefont {P.~F.}\ \bibnamefont
  {Zweifel}},\ }\href {\doibase 10.1063/1.1705323} {\bibfield  {journal}
  {\bibinfo  {journal} {J. Math. Phys.}\ }\textbf {\bibinfo {volume} {8}},\
  \bibinfo {pages} {1097} (\bibinfo {year} {1967})}\BibitemShut {NoStop}%
\bibitem [{Note1()}]{Note1}%
  \BibitemOpen
  \bibinfo {note} {A more physically appealing derivation proceeds by scaling
  the equation so that an expansion of the Liouville operator in small
  parameter $(m/M)^{1/2}$, where m and M are the characteristic masses of the
  quantum subsystem and bath DOF, can be carried out ( see Ref~\protect
  \rev@citealpnum {Kapral:1999/jcp/8919}).}\BibitemShut {Stop}%
\bibitem [{\citenamefont {Kernan}, \citenamefont {Ciccotti},\ and\
  \citenamefont {Kapral}(2008)}]{Kernan:2008ge}%
  \BibitemOpen
  \bibfield  {author} {\bibinfo {author} {\bibfnamefont {D.~M.}\ \bibnamefont
  {Kernan}}, \bibinfo {author} {\bibfnamefont {G.}~\bibnamefont {Ciccotti}}, \
  and\ \bibinfo {author} {\bibfnamefont {R.}~\bibnamefont {Kapral}},\
  }\href@noop {} {\bibfield  {journal} {\bibinfo  {journal} {The Journal of
  Physical Chemistry B}\ }\textbf {\bibinfo {volume} {112}},\ \bibinfo {pages}
  {424} (\bibinfo {year} {2008})}\BibitemShut {NoStop}%
\bibitem [{\citenamefont {Ferretti}\ \emph {et~al.}(1996)\citenamefont
  {Ferretti}, \citenamefont {Granucci}, \citenamefont {Lami}, \citenamefont
  {Persico},\ and\ \citenamefont {Villani}}]{Ferretti:1996/jcp/5517}%
  \BibitemOpen
  \bibfield  {author} {\bibinfo {author} {\bibfnamefont {A.}~\bibnamefont
  {Ferretti}}, \bibinfo {author} {\bibfnamefont {G.}~\bibnamefont {Granucci}},
  \bibinfo {author} {\bibfnamefont {A.}~\bibnamefont {Lami}}, \bibinfo {author}
  {\bibfnamefont {M.}~\bibnamefont {Persico}}, \ and\ \bibinfo {author}
  {\bibfnamefont {G.}~\bibnamefont {Villani}},\ }\href {\doibase
  10.1063/1.471791} {\bibfield  {journal} {\bibinfo  {journal} {J. Chem.
  Phys.}\ }\textbf {\bibinfo {volume} {104}},\ \bibinfo {pages} {5517}
  (\bibinfo {year} {1996})}\BibitemShut {NoStop}%
\end{thebibliography}

%

\end{document}